# Cyberattacks on Energy Infrastructures: Modern War Weapons

Tawfiq M. Aljohani
*Department of Electrical and Computer Engineering, Taibah University (Y-Campus), Saudi Arabia 42353*
torwi@taibahu.edu.sa

*Abstract-* **Recent high-profile cyberattacks on energy infrastructures, such as the security breach of the Colonial Pipeline in 2021 and attacks that have disrupted Ukraine's power grid from the mid-2010s till date, have pushed cybersecurity as a top priority. As political tensions have escalated in Europe this year, concerns about critical infrastructure security have increased. Operators in the industrial sector face new cybersecurity threats that increase the risk of disruptions in services, property damages, and environmental harm. Amid rising geopolitical tensions, industrial companies, with their network-connected systems, are now considered major targets for adversaries to advance political, social, or military agendas. Moreover, the recent Russian-Ukrainian conflict has set the alarm worldwide about the danger of targeting energy grids via cyberattacks. Attack methodologies, techniques, and procedures used successfully to hack energy grids in Ukraine can be used elsewhere. This work aims to present a thorough analysis of the cybersecurity of the energy infrastructure amid the increased rise of cyberwars. The article navigates through the recent history of energy-related cyberattacks and their reasoning, discusses the grid's vulnerability, and makes a precautionary argument for securing the grids against them.**

*Index Terms—* Cybersecurity of Critical Infrastructure, Russia-Ukraine war, Smart grid hacking, cybersecurity of power grids, power system digitalization, energy cyberwar.

## 1. Introduction

For years, experts have expressed growing concerns that critical infrastructures are vulnerable to cyberattacks almost anywhere in the world. Remotely transmitted malware and viruses can compromise interconnected power systems, airports, hospitals, military systems, and anywhere internet-connected computers are used to manage critical functions. As the world heads toward more digitalization, integrating new communication technologies offers unlimited possibilities for the proper accumulation of information, creating a fundamental dependence on proper functioning in all areas of society. As such reliance could offer a great opportunity for a smart and efficient world, it also poses significant threats to the operation of critical infrastructures and vital facilities. Specifically, safety issues from illegal acts that violate the integrity, availability, and confidentiality of information in cyberspace that could damage information resources and communication systems are expected to be the central theme in the digitalized world. Exploitation of cyberspace to advance specific political, social, and military agendas is of particular concern.

In recent years, data and information resources from industrial control systems (ICS) in energy utilities, financial institutions, government agencies and official servers have become the main target of frequent cyberattacks. The sharp rise in cyberattacks shows that politically motivated cybercriminal groups, primarily acting on behalf of state governments, have carried out proliferating and infringing attacks on public and private entities, resulting in severe repetitional losses. For example, Russia, known for its advanced cyber-capabilities, is seen as the main villain in cyberspace due to accusations of cyberattacks against its adversaries to promote political and military objectives. On April 20, 2022, cybersecurity authorities from the United States, Canada, Australia, New Zealand, and the United Kingdom issued a joint advisory, warning organizations that Russia's invasion of Ukraine could evoke malicious cyber threats worldwide [1]. The adversary activities are due in response to the unprecedented economic sanctions imposed on Russia and military assistance provided by the U.S. and its allies to Ukraine to help it withstand the Russian invasion.

Energy cybersecurity has become under the spotlight with the unfolding of Ukraine's war crisis, even though concerns about it have been evident long before the crisis itself. The swift penetration of renewables and the digitalization of energy networks have further increased the vulnerability of the energy systems. A future that depends on low-carbon technologies involves electrifying transportation and industrial processes. Sectors of the economy that were initially fueled by fossil fuels are currently being connected to energy systems that are under digital control, making them easily accessible to hackers. Additionally, a future with net-zero carbon emission further implies that a more decentralized generation system will be in use. In contrast to using a few high-capacity coal- or gas-fired power stations to supply energy, a series of dispersed solar panels and wind turbines will be adopted. These facilities significantly increase the surface area of the energy system exposed to attacks with the inclusion of the extensive communication technologies that couple the power cables, substations, and storage units to achieve swift and smart control.

The nature of broad-scale cyber and electronic warfare attacks is yet to be fully understood. The continued Russian-Ukrainian conflict is an example of the rise of cyber threats and cyber activists that target critical infrastructures, especially energy systems. Russia acutely disrupted the power grid of Ukraine in 2015, 2016, and with the subsequent unleashing of the destructive NotPetya malware attack in 2017 which had global effects and resulted in billions of dollars worth of damage. Ever since its first invasion of Ukraine in 2014, Russia has exploited the former Soviet Union state like a playground to test its emerging cyber weapons. A comprehension of Ukraine's grid structure and technological potential is possible with its historical Soviet infrastructure that could have played a role in the successful implementation of hundreds of cyberattacks that, with its major share targeting the energy infrastructure, aimed to collect intelligence on the Ukrainian forces and critical bases. Consequently, the use of aggressive cyber and electronic warfare in the pre-and first phases of the 2022 war could have played a major role in Russia's decision to invade Ukraine, possibly giving them advantages on the ground. According to an MIT technology review report [2], massive cyberattack campaigns were carried out on the Ukrainian energy utilities during the first hours of the recent military invasion. Although effect of such attcaks on the military operations were not yet assessed, it was announced that the cyberattacks aimed to delete data from the assailed utility control rooms. Before the invasion, the defacements of the Ukrainian websites, destructive malware, and distributed denial of service (DDoS) attacks were believed to have been caused by Russia and its delegations. The Ukrainian cyber response teams identified and halted Russian military hackers attempting to penetrate power substations in Ukraine which could have resulted in 2 million people losing power [3]. On the other hand, the "IT Army," a volunteer group consisting of various hacktivists, has been employed with the responsibility to initiate DDoS attacks on the Russian and Belarusian governments, including their energy and banking systems, as well as the task to identify and report Russian disinformation campaigns. For instance, the Belarusian Cyber Partisans have paid more attention to inflicting mayhem on the Belarusian train system to keep the transition of Russian troops and equipment in slow motion [4]. The additional announcement came from the decentralized hacking group 'Anonymous' that they are officially involved in a cyberwar against the Russian government. Its main targets are state-run power grids, radio and television, and communication networks. This has also ignited calls from the other side, where criminal groups in allegiance to Russia have publicly declared their support for the recent war and promised retaliative actions against the infrastructures of nations who might attack Russia [5]. It is worth mentioning that Ukrainian authorities, by August 2022, have documented over 300 ferocious cyberattacks that hit vital points around the country per the CyberPeace Institute.

The major contribution of this work is to provide a discussion on the cybersecurity of the energy infrastructure. Specifically, the paper discusses the logic behind targeting power grids in recent years with different types of malware; provides an insight into the types and techniques used in cyberattacks and whether the power grids are vulnerable to such attacks; discusses the motives behind initiating cyberattacks and standardization efforts made as counterefforts; and finally concludes with remarks on the resiliency of the grids to achieve improved security.

**2. From Airstrikes to Cyberattacks: Why Targeting the Energy Infrastructure During Conflicts?**
*2.1 Attacking power grid during wars.*

The energy infrastructure is the backbone of any society and is especially important as it facilitates the operations of other critical infrastructure sectors. There is a pearl of conventional wisdom about targeting power networks that diversity of institutions can be broadly affected by such attacks, from political, social, and military aspects. Particularly, two main political consequences are the outcome of electricity losses. First, there will be a weakening of civilian morale and, therefore, forcibly alter the behavior of the assailed governments. Secondly, these attacks can aggravate the expenses being made by a country's political leaders, compelling them to adjust their policies. Besides, two main critical military consequences are frequently pointed out: the loss of power will directly affect the resisting forces and likely reduce the assailed country's capability of sufficient war equipment production. Whether separately or in combination, these four arguments were employed long years ago to support launching attacks on the enemy's electric systems. The discussion on such impacts is illustrated in section 3 of this work.

It is believed that the fastest and most effective way to initiate a direct attack on the will of a nation is "*to cripple its economic structure and intimidate its very existence*" through precision attacks [6]. Due to many possible strategic targets within an economic structure, it is difficult to hit every target, even with a massive air force. However, experts at the Air Corps Tactical School (ACTS) highlight that a developed nation relies heavily on its energy grid to support various vital points, making it the best strategic target for any attack. Moreover, the experts in ACTS believe that victory in any war heavily relies on the civilian population's will to remain persistent in fighting; hence, their target analysis highlighted civilian rather than military impacts [7]. This highlights choosing power networks as strategic targets during conflicts. Certainly, the outcomes of these attacks would not only impact war production but also influence the will of civilians. The morale of the civilian population could be affected by attacks on power grids to the extent that it likely will induce modifications in government policy. The breaking up of the economic structure through the destruction of power grids could be an efficient way to execute such a strategy. These theories about strategic attacks created at the ACTS became practical concepts that greatly inspired the targets during wars such as Vietnam and Iraq's first and second wars.

In addition to being a strategically important target, attacks on energy infrastructures using newly developed and possibly untraceable techniques, like cyberattacks, have become widely popular in recent years. Asymmetric warfare is a technique used in wars for hundreds of years. Whenever there is an unequal fight between contenders, the weakest fighters would use any new, untraditional, less-costly methods to harm the strongest fighter. Modern forms of asymmetric warfare involve the use of computers. According to Clark and Knack, electronic warfare is the intrusion of foreign computers or networks to disable them in order to neutralize and disrupt essential services [8]. Hence, exploiting the internet and network systems provides weak countries a golden opportunity to fight traditionally stronger enemies through doing more extensive espionage. The U.S. is the most powerful country in terms of conventional military power. It spends far more on defense than any other country, such as China or Russia, which are the biggest threat to the U.S. in terms of geopolitical and conventional military power. China and Russia have developed significant offensive cyber capabilities to complement deficiencies in their military, and see the use of the U.S. power grid as a strategic target during wartime [9]. Historically, the two countries have used cyberspace for espionage. While China is paying more attention to stealing U.S. research to advance its project programs and identify U.S. military strategies and weaknesses that could be exploited in wartime, Russia poses a more direct threat to the U.S. energy grid and critical infrastructure sectors. Russia is increasingly focusing on devising offensive cyber capabilities, including cyber weapons. This is evidenced by attacks on the French TV5-Monde and the Ukrainian energy system. Furthermore, these attacks have demonstrated cyber capabilities to disrupt physical infrastructure [10]. The TV5-Monde attack used specialized malware to target and destroy the encoding system to broadcast the French TV network. On the other hand, the cyberattack on Ukraine's power system used a variant of the BlackEnergy malware and a modified kill-disk function. The disk-remove function erased the master boot records (MBR), rendering the component obsolete. Disabling the MBR disconnected 30 substation breakers using the human-machine interface (HMI) [10, 11].

The interest of Russia in meddling with the U.S. energy infrastructure is not a secret and has been brought to light recently. A report from David Sanger, a New York Times journalist with a cybersecurity background, reveals that '*According to the United States intelligence officials and technology company executives, Russian hackers whom the State sponsors seem to focus more on indicating that their technology can disrupt the American electric utility grid compared to the events surrounding the November 2018 elections*.' The same article also highlights Sanger's report that the Russian military intelligence agency had penetrated the control rooms of power plants across the United States according to Department of Homeland Security (DHS) and that there are fears that there are parts of the grid could be hijacked by remote control. The DHS well understands the scope of the threat, says Sanger. However, he also unveils that great fear regarding the fact that plans, by the Russians, may be underway to disconnect American power systems in moments of conflict and which induce a military response. According to U.S. federal officials, Russian hackers reported in 2017 that hundreds of victims were claimed during a vast and persistent cyberattacks campaign that gave them access to the inner chambers of the control rooms of U.S. electric utilities that led to widespread blackouts. The DHS Chief of Industrial Control System Analysis, Jonathan Homer, revealed that there were cases of cyberattacks in the spring of 2016 into 2017 and that such attacks took advantage of connections between utilities and vendors who had unique access to how software could be updated, how diagnostics on equipment were run and how specific services were performed to ensure millions of pieces of gear were kept in working order [12]. Indeed, a successful cyberattack on the power grid could be devastating to economies and human lives. On August 14, 2003, eight states in the U.S. and Canada experienced a major blackout. The outage has nothing to do with cybersecurity, but it could provide insight into the impact of a successful attack on the U.S. grid. About 50 million people lost energy for two days; the blackout also killed 11 people and cost the U.S. economy about $ 6 billion [13].

*2.2 Vulnerability of Power Grids Against Cyberattacks: Is The Power Grid Hackable?*

Power grids are vulnerable to cyberattacks. Before automation took place in recent years, SCADA systems were not initially connected to the internet and hence did not receive significant threats from external attacks, as most concerns were from inside threats (e.g., misoperation). Connecting SCADA to the internet presents security concerns due to weak open network protocols such as TCP/IP that are more cost-effective than other proprietary methods. As more SCADA systems are deployed, open network protocols allow businesses to connect SCADA systems, which pave the way to both a more digitalized and intelligent energy infrastructure and a more vulnerable and jeopardized one. Additionally, the use of weak communication protocols such as Modbus and DNP3 creates more security concerns. Modbus and DNP3 lack proper authentication and encryption, allowing anyone to view transmitted data over the network in plain text. Data transmitted via Modbus or DNP3 can also be intercepted, manipulated, or altered [9]. Hackers view vulnerabilities in ICS infrastructure as the holy grail of cyber warfare [14]. An attacker can compromise ICS and place it outside its normal operating range. Stuxnet, for example, increased the speed of centrifuges at Natanz, a uranium enrichment plant in Iran. The standard operating range of the centrifuge at Natanz was 63,000 revolutions per minute (RPM). The malware raised the speed to 84,600 rpm and lowered it to 120 rpm, disrupting the centrifuge by changing its rotational speed [15].

Companies can control internet-connected SCADA systems hundreds of miles away, where employees do not have to be physically present at the stations. As more SCADA systems are integrated, the need for connectivity increases. Thousands of new devices such as sensors, controllers, relays, and meters added to the smart grid pose security concerns for a variety of reasons. The more devices connected to the smart grid, the more it gives attackers a broader attack surface. While vendors of these technologies have standardized them to provide better interoperability, they unintentionally created security holes. In some cases, vendors install backdoors on devices to apply patches or software updates. Other vendors void device warranties if factory settings are changed, such as changing factory passwords or installing software that provides greater security [9]. While remote access has increased the vulnerability of SCADA systems, remote access is essential to managing the distributed assets inherent in the power system. Strong passwords, authentication, and encryption are often overlooked when implementing remote access tools. Authentication verifies a user's identity, typically using a username and password. On the other hand, encryption makes data unreadable for anyone other than authorized users. The lack of strong passwords, authentication, and encryption is more likely to compromise the utility's control system network. Tools like SHODAN allow a person to search for internet-accessible control systems, routers, and other internet-accessible devices. Hence, fears arise that hackers can use similar tools to search for various internet-connected equipment and to identify devices by their IP addresses [16]. Vulnerabilities in SCADA and other ICS systems continue to grow every year primarily due to the fact that ICS components have become increasingly accessible over the internet. According to Kaspersky Lab researchers, the number of vulnerabilities discovered in ICS components in power utilities has increased from 19 in 2010 to 189 in 2015 [17]. The US Cybersecurity and Infrastructure Security Agency (CISA) announced that this number reached 600 security disclosures in the first half of 2022. Security experts argue that most of these vulnerabilities are yet to be resolved, with a significant share of them remaining unpatched, meaning that they have not yet received an update to address them. Fig. 1 presents an illustration of conducting cyberattacks on an energy control center.

*2.3 History of cyberattacks on energy infrastructures*

The most infamous cyberattack on energy systems happened in 2015 when hackers cut off power to almost a quarter-million people in Ukraine for nearly six hours. The attack, credited to Russia-backed hackers, was successful due to the lack of proper isolation between the Operational Technology (OT) and Information Technology (IT) systems. Hackers disrupted the I.T. systems through successful phishing email attacks that granted them free movement throughout the network to assault the utility's energy management system. The attackers implemented firmware that affected the ability of the grid operators to communicate with substations while also influencing the operation of the station's appartus. The following year, another cyberattack was successfully launched through a substation north of Kyiv, knocking out every circuit breaker in the system and briefly shutting down power in large parts of the city. According to Andy Bochman, a senior cybersecurity strategist at the Idaho National Laboratory, the goal behind these attacks were not only to punish the Ukrainians but also to show what Russia can do to other adversaries. Another attack, called the Notpetya, took place in 2017 and was successfully executed by a Russian military intelligence group resulting in the infection of 10% of all Ukrainian computer systems with a malware package which ended up spreading globally. Estimated losses in banking, military and energy industries are nearly $10 billion due to the attack which was considered one of the most severe cyberattacks in history. The authors of [18] suggest that the U.S. power grid could be hacked with the same techniques used to hack the Ukrainian grid.

To a great extent, grid impacts from cybersecurity threats have been noticed in recent years in the U.S. and worldwide. A 2018 attack disrupted communications on the Midcontinent Independent System Operator (MISO) grid without significantly impacting customers. With the advent of the outages due to such attacks in Ukraine, experts believe the U.S. grid is impermeable to how hackers can launch cyberattacks, triggering calls to address the issue. In 2007, it was demonstrated by the Idaho National Laboratory that a cyberattack could physically crush a generator by placing it in an out-of-phase mode, resulting in an extreme torque and breakdown of the machine. Similarly, the North American Electric Reliability Corp. (NERC) also revealed that a quarter of the electric utilities it control were exposed to the vulnerability of the 2019 SolarWinds attack, which is characterized by the insertion of advanced malware into the software supply chain [19]. Experts say that it could take several years for the electric sectors to estimate the full impact of that attack. Similarly, while there was no impact on the grid resulting from the attack on Colonial Pipeline, a transport pathway for refined oil products in Southern U.S. States, the attack serves as an example of undesired consequences. The attack on Colonial's I.T. system prompted a defensive shutdown by the company, leading to hefty costs. Similarly, SolarWinds exemplifies the threat to supply chains, where a single platform was compromised and afterward infected thousands of users. The Colonial shutdown reveals that hackers do not essentially need to target control systems to instigate societal impacts. Anthony Pugliese, a high-ranking official at the U.S. Federal Energy Regulatory Commission (FERC), stated that there was an increase in the number of oppositional countries that paid attention to pipelines as areas of enormous opportunity. Such a statement becomes evident as a wave of cyberattacks aimed to target the communication systems of five U.S. pipeline firms since 2018. Fig.2 provides a timeline of the most notable cyberattacks that targeted energy infrastructures in recent years.

## 3. Reasoning Behaind Attacking the Power Grids in Modern Times

Experts at ACTS suggest that to successfully conduct tactical attacks against energy networks, the economic structure of a country needs to be well-understood. Information is required to initiate successful attacks on energy systems, such as level of consumption, backup systems, and the predictable impacts of eliminating electricity. As a result, collection of such data justifies performing another set of cyberattacks on other entities with a mission to collect sufficient intelligence. An understanding of the tactics behind the attacks could be more consequential than deciding on the grid's vulnerability. This is because the timing for effective attacks on such facilities greatly depends on the strategies to be used. As mentioned in section 2, four basic techniques have been long applied, either discretely or in combination, to justify adversarial attacks on electric power, whether this attack is performed by airstrikes or cyberattacks.

*3.1 Attacks to target citizens' morale*

The assumption that cutting off the energy supply to civilians will induce a modification in a nation's policy has been one of the most persistent beliefs in tactical wars, as seen in ACTS strategic targeting policy [7]. In the same manner, intermittent elimination of electricity via cyberattacks may produce a decline in civilian morale and, if successfully implemented, could force government policies to change their intended goals. One of the strategies employed in the recent Russian-Ukrainian conflict seems to aim to influence the citizens' morale. During the first hours of the military operations, major cyberattacks hit energy substations across the eastern frontier of Ukraine. While some aimed to delete control rooms data, vital for proper grid operation, other targeted circuit breakers and heat and power stations located at Sumy, leading to outages and wide services disconnection on citizens in the cold month of February [20].

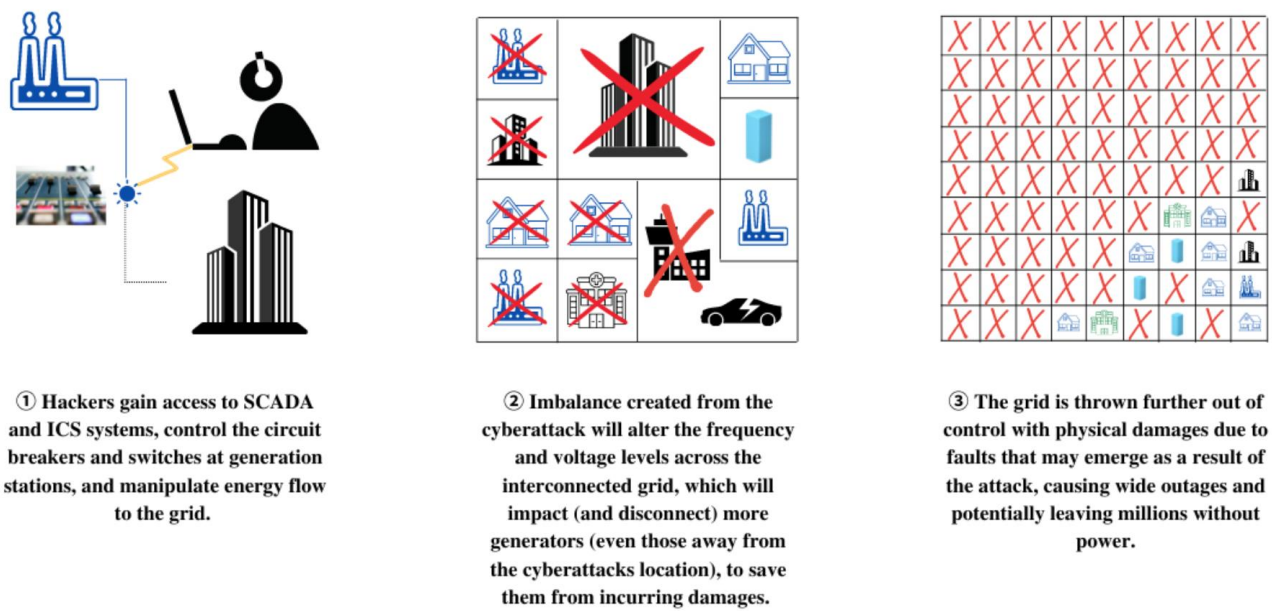

Fig.1 Illustration of possible outcomes of hacking utility control systems

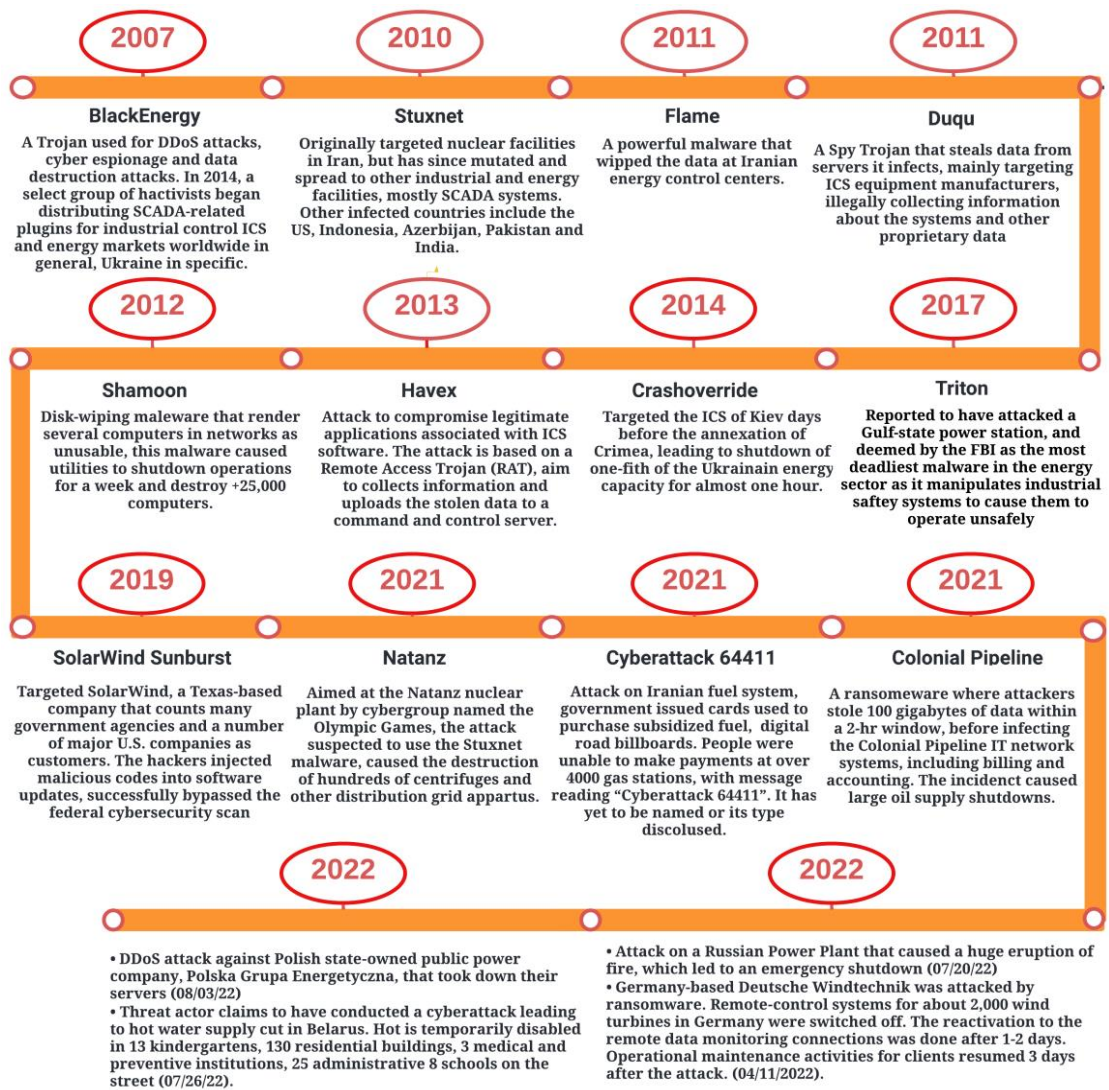

Fig.2. Timeline for the major cyberattacks against energy infrastructures in recent years

Additionally, there is another way of indirectly influencing the citizens' morale via disabling critical pipelines or gas processing facilities to cause a considerable fuel supply disruption. The fuel disruption will likely lead to a significant increase in fuel prices, which becomes a tool to influence the leaders by igniting inconvenience and anger among the citizens. While such a scenario occurred with the Colonial pipeline cyberattack in 2021, the direct effect on the comfort of citizens and surcharges in energy prices are still underexplored.

*3.2 Attacks to influence leaders' and policies*

Attacking a vital infrastructure like the energy systems has been one of the strategic moves to manipulate political leaders' decisions. Whether attacks aimed to cut power on civilians or destroy expensive equipment at the energy grids, such attacks escalate costs on leadership which could drive a policy change. However, such strategy may not be completely effective due to numerous reasons. First, a potential cause is the sense of nationalism and the high hopes most nations possess during a conflict, which likely despises the typical cost estimates versus benefits that should apply to external nations not involved in the battles. The damage wreaked on electric power will possibly not go beyond the costs that the leaders of a country desire to pay if the problem in question is of great national interest. Secondly, immediately after national leaders commit themselves to a course of action, they become unwilling to bend their minds. Such a change of decisions could imply that they would lose their reputation and political power. Instead of conceding to some defeat in domestic politics, they prefer to proceed with their current plan of action. In a more practical sense, political leaders are left out of the consequences of losing the national power system. While there are yet to have studies that could assess the impact of cyberattacks on political leaders' decisions, an official USAF bombing survey that aimed to evaluate the effect of taking out energy service via airstrikes concluded that any attack campaigns launched at a country's electrical grid are unlikely to alter political authorities positions [7].

*3.3 Attacks to gain military advances*

Targeting energy networks to alter a nation's military forces appears to be prevalent in recent conflicts. This is mainly a manifestation of how seriously reliant the military is on electricity to execute activities like powering air defense radars and communication systems compared to the past. During the first weeks of the 2022 invasion, Russia surrounded and successfully conquered the Zaporizhzhia nuclear power plant that provides the energy to most of the Eastern part of Ukraine, apparently where most of the Russian military operations are conducted. Although cyberattacks on power plants could serve as a strategic measure to generate brief confusion, as happened during the first hours of the Russian invasion of Ukraine when widespread cyberattacks targeted control centers in Eastern Ukraine, their impact could be of minimal long-term consequences. This is attributed to the fact that the military, as a priority user, will gain access to any available power in the national grid and have full authority to possess emergency power systems [21]. Doubtless, confusion was induced in the Iraqi military due to the airstrikes on the Iraqi power system in 1991 and 2003. Yet, its impact in advancing the objective of tactically crippling the Iraqi leadership is neither well-understood nor discussed. After an investigation into the bomb wrecks in Iraq aftermath of the 1991 war, William C. Arkin of Greenpeace International believed that the tactical bombing had almost no difference in the outcome [6]. Moreover, the target selection plan during the Gulf Wars has been questioned due to the attack on the power grid and the implicit collateral wreckage it caused. Experts at The Gulf Conflict and the New World Order, Lawrence Freedman and Efraim Karsh reveal that "*The areas of USAF campaign directly aimed at the economic and political structure of the Iraqi system appear to have had little relevance to the final victory*." However, this may not be the case for cyberattacks since airstrikes campaigns are associated with significant human, economic and military costs. On the contrary, cyberattacks could be discretely performed hundreds of miles away and at high precision to achieve the overall goals at significantly lower costs than air campaigns. This may justify using cyberattacks as a strategic tool to mangle a nation's energy infrastructure during conflicts since potential military gains could be achieved with minimal physical efforts. Among the cyberattacks that hit numerous energy control centers during this war, the Ukrainian government announced in April that they successfully foiled a Russian malware aimed at a regional power utility. The malware, similar to the one that took down most of Kyiv's power grid in 2015, was said to be successfully planted in February. The Ukrainian officials believe that Russia waited until April to perform the attack as a part of its military advancing strategy [20].

Another facet of using cyberattacks to achieve military gains is data collection to build sufficient intelligence about a country. During the months that preceded the recent Russian invasion, hundreds of cyberattacks targeted different critical facilities in Ukraine, aimed primarily at collecting intelligence. Among the targeted infrastructure, energy centers had a significant share, where malware believed to be built and executed by Russian-backed cybergroups flooded Ukrainian and other European energy control centers. The collected information may have played a role in the decision to invade Ukraine, as the concentration of energy generation and consumption levels provide hints about the vital military, economic and industrial facilities of the assailed country.

In addition, undermining the energy grid could indirectly advance other military objectives. One of the military tactics that hackers could serve is to intentionally create disputes between nations on the assumption that one has cyberattacked the other. Such a scenario could be achieved by initiating cyberattacks with attributions that fell on other countries. For example, Russia is suspected of being behind attacks on TV5-Monde; however, attribution and verification issues were disputed. Russia appears to have carried out a false flag operation during the TV-5Monde cyberattack to trick investigators that a cyberterrorist was behind the attack. APT28, a Russian cybercriminal group, hacked TV5-Monde's Facebook account and posted pro-ISIS propaganda and threats to forces involved in operations against ISIS. Such tactics could be effective for countries interested in attacking the U.S. power grid without declaring full responsibility for it. A country can use known Russian servers, methods, tactics, and procedures to give the impression that Russia carried out the cyberattack. Hence, retaliation by the targeted countries after a cyberattack could be very low due to attribution. Attribution provides the government with the possibility of plausible denial, as the assailed country must prove that the attacking country sponsored the cyber espionage.

As evidenced by the fact that Russia added cyber operations to its wartime doctrine, Russia uses cyber operations to enhance its military operations. The 2008 attack on Georgia is a testament to the state's use of cyberattacks to support conventional military attacks. Russian hacktivists organized large-scale DDoS attacks against Georgian media and government websites to indicate Russia's powerfulness and capabilities against its adversaries.

*3.4 Attacks on war equipment production*

The stoppage of (or delay in) war production has been the most powerful argument for attacking energy systems. War goods-making industries rely heavily on electricity, and many processes are impossible without this resource. The bulk of electricity generated in most countries is utilized in the manufacturing process. A conclusion was reached in [7] following World War II that attacks on energy systems can be central in achieving a long-term win in a war against a country lacking the ability to import. Similarly, in short-term war, where the enemy has amassed war items, halting war production could also result in a sufficient impact on winning the war. Consequently, sponsoring cyberattacks on electrical grids to impact war production or even collect data on military objectives could strategically be a highly efficient move. It is worth mentioning that there have yet to be sufficient reports to assess the impact of cyberattacks on Ukraine, enumerated to be over 300 attacks until August 2022, on its war production or manufacturing capabilities.

**4. Gaps, Standardization and Mitigation Efforts**
*4.1 Gaps in securing energy grids cyberly*

The smart grid is a cyber-physical system comprising different layers representing the attack surface, including physical, sensor/actuator, network, control, and information. With increased internet connectivity, highly confidential personal details are readily accessible. More expansive attack surfaces are created for various kinds of cyberattacks on the grid due to their connection to the internet. DDoS, False Data Injection Attacks (FDIA), and the use of malware are well-known tools against the grid. While hardware failures or software crashes are common problems in the operation of the grid, they are categorized as general system failures with no external causes. Similarly, misoperation caused by external services is usually defined as third-party failures (e.g., internet service providers). Regarding the cybersecurity of the grid, the external or third-party failures are irrelevant. What is important is the security loopholes that hackers see as gates to intrude the grid. These loopholes could be weak points like domestic appliances that could be hacked, especially if defensive mechanisms do not have a solid foundation from the consumer perspective and I.T. systems are not properly segmented. This could provide an access point to the broader energy system. Another loophole is electric vehicles (EVs), which are expected to be millions on roads in the coming few years. A directed attack on EVs and charging stations could trigger substantial disruption to the local power supply. In addition, high-wattage internet-of-things devices such as air conditioners and heaters could be utilized to begin large-scale organized attacks on the grid, resulting in local power outages.

The cybersecurity deficiencies within the power grid, which primarily originate from the development of the O.T. and I.T. systems, were reviewed in [22]. Using antiquated cyber defenses by operators of critical infrastructures may further increase the vulnerabilities of the power grid and make it less complicated for hackers to launch attacks. While the National Infrastructure Protection Plan (NIPP) is created for critical infrastructure operators to adopt; however, it is quite challenging to implement some of them. The authors of [23] utilized a grounded theory approach to underline the insufficiency of cybersecurity strategies used by the U.S. government against increasing cyber threats, highlighting gaps in national policy. The relatively high dependence of critical infrastructure on ICS and SCADA systems will lead to massive cyberattacks not only on state or private operators but also on the end-users. As a result, preventive mechanisms must be integrated sufficiently to ensure the safety and security of critical infrastructures like the energy system and its connected users. Such mechanisms include removal of network vulnerability as well as the access to the internet of all ICS; exclusion of ICS networks and devices from the enterprise network; establishment of firewalls for ICS networks and devices; identification of devices that have remote access, and the integration of a secure virtual private network (VPN) to gain access to component(s); a removal or modification of default system accounts and passwords, implementation of a solid password policy; identification and recognition of vulnerabilities; supervision of the administrator accounts created by third-party vendors, and regular performance of audits that recognize cybersecurity risk and impact.

*4.2 Standardization*

The purpose of standardization in cybersecurity is to improve the security of I.T., O.T., ICS systems, networks, and critical infrastructure. According to NIST, cybersecurity standards define functional and safety requirements within a product, design, process, or technology environment. Well-developed cybersecurity standards enable consistency among product developers and serve as a reliable guide for purchasing security products. The standards cover a wide range of granularities, from mathematical definitions of cryptographic algorithms to specifications of web browser security features, and are typically implementation-agnostic. In reality, the standards must address user needs and be practical, as cost and technical limitations must be incorporated when manufacturing products to meet standardization goals. In the U.S,, for example, the major cybersecurity collaboration between the energy industry, the Department of Energy (DOE), and the DOE Office of electricity delivery and energy reliability (DOE-OE) is through the Cybersecurity Risk Information-Sharing Program (CRISP). Through CRISP, quality information regarding cybersecurity can be regularly exchanged among various actors within a defined structure. Also, industrial operators have played a key role in identifying cyber threats and accumulating protection standards through a partnership with the U.S. Federal government, namely the Information Sharing and Analysis Centers (ISACs). A specialized ISAC also exists for the electricity sector (E-ISAC), whose primary function is to act as the middle link between the DOE and utilities to coordinate and exchange vital information. Table 1 shows the most recent standardization efforts, both in the U.S. and Europe, to improve the cybersecurity practices related to the energy industry.

*4.3 Suggested mitigation efforts*
*4.3.1 Improving the resiliency of the energy infrastructure*

The power grid still faces many threats from cyberspace. Smart grid vulnerabilities will always be there as the number of proactive hackers with sufficient tools is expected to explode in the coming years, along with new ways to espionage their targets. However, the chances of exploitation can be reduced by implementing cyber resiliency. One way to improve cyber resilience is to optimally design and implement security mechanisms in devices integrated into the smart grid. Many electronic devices are designed and incorporated without implementing clear security measures, which will always present significant security concerns for potential hacking. Therefore, the network must increase its resilience to cyberattacks. The idea of cyber resilience is to make the system more difficult to access and consequently unattractive to threat actors because it would be too costly and time-consuming to exploit. In-depth defenses is the cornerstone of cyber resilience as it provides several layers of protection to prevent a successful cyberattack. Additionally, redundancy is essential to ensure that in case of the failure of a security measure, others are activated to maintain the network's security. A solid resilience plan can be created by combining techniques presented in fig. 3 to offer uncompromising protection against potential cyberattacks.

*4.3.2 Acceleration of proper cyber-breach notification system*

Hackers often have an advantage over grid operators who do not report cyber incidents promptly, which severely impact proper mitigation efforts. Timely cyber-breach notices activate the obligatory disclosure of information among appropriate parties, including the nature and scope of cyber espionage, leading to better handling of the situation. Such procedure should be accelerated by the government and private sectors as quickly as possible, knowing that it takes time to pass laws while the cyber threat scope continues to grow and become an increasing emergency.

*4.3.3 Accelerate the application of commercial and open-source intelligence (OSINT).*

OSINT refers to the method of obtaining information from public sources. Most intelligence professionals extend this definition to include information intended for public use that anyone can find over internet searches, public announcements, mass or social media,...etc. The use of OSINT does not mandate its advocates to hack into systems or use confidential credentials to access data. In ethical hacking, OSINT could be an effective tool to uncover the digital footprints of cyberattacks on the energy infrastructure, eventually leading to a better understanding of the performed espionage. Nonetheless, a drawback could emerge as using sensitive public information may allow malicious hackers to use it as assets to launch future cyberattacks creatively. Therefore, tradecraft regulations that elucidate the application of OSINT in analytic products need to be well established and maintained.

**Steps to Improve the Resiliency of the Energy Grid**

**Improved Firewalls**
Security measures that are employed in securing ICS and SCADA networks. They must have the capacity to play out different functions especially the prevention of external threats. A key feature of firewalls is the deep packet inspection which is applied to control system networks and executes more in-depth probation into network traffic, which could be exploited in distinguishing malicious network traffic from regular types.

**Full System Hardening**
This technique ensures that every unecessary application or service are removed, improving the security of the system by imposing constraints on the dimension of surface area that could be occupied by a threat actor to initiate an attack. To achieve system hardening, separation of services is executed and this makes sure that cross-contaminations are mitigated especially in the case that one device is compromised.

**Application Whitelisting**
A well-known security solution for embedded devices that serve unique functions within a larger system. Rather than block bad applications, application whitelisting only permits recognized applications. While the number of *good* applications remains finite, that of *bad* applications is relatively infinite. Application whitelisting can be applied to SCADA servers as it is effective against memory attacks such as buffer overflows. The combined usage of the application whitelisting with anti-virus software can be employed as a mechanism for auditing.

**HIDS/HIPS**
Called the host-based intrusion detection system, this software is installed as a monitoring tool in probing log files, modifications to system configuration and illegitimate access by authorized users to delicate data. By inducing prompt warning signals against an attempt being made to illegitimately access the system, the HIDS mitigates any intrusive activity. This mechanism also applies to the HIPS which cross-examines and dynamically counters the activity of the intruder.

**Encryption of Communication Data**
This technique, as an essential part of cyber hygiene, is applied to disallow illegitimate users from gaining access to communication data. IEEE P1711, a tool that defines a cryptographic protocol for the cybersecurity of substation serial links, is widely used in power system communication. Encryption between SCADA devices is advised by NIST to protect communication and prevent the interception of data by illegitimate users.

Fig.3. suggested actios for a resilient energy grids against cyberattacks

*4.3.4 Mandate proper interagency parternship to optimally identify actions, capabilities, and threats.*

The primary benefit of enabling a proactive, periodic interagency collaboration within the energy industry lies in safeguarding the confidentiality of energy-related private and public data; the availability and integrity of energy-related technologies; and strengthening the cybersecurity of involved stakeholders. Small to medium energy service providers, which usually form a significant part of the whole energy systems, frequently lack the sufficient resources to invest in complete cybersecurity protections. In contrast, large energy providers and grid operators are expected to operate under strict federal- or state-mandated regulations to protect their infrastructure. Therefore, it is necessary to cautiously examine grid vulnerabilities and develop strategies to tackle such inadequacies that may emerge due to the absence of interagency partnership. At least, local commissions should place demands on smaller utilities to actively incorporate security protection measures similar to those implemented by large utilities.

*4.3.5 Encourage public and private sector parternship to upgrade cyber training, research and development.*

Criticisms have been heaped on major energy utilities for their lack of expenditure on cybersecurity, given that there are rising threats in this regard. It would be a rule of thumb to note that energy utilities should channel enormous resources towards improving cybersecurity to avoid hefty costs if attacked. This is against the reality of the current scene, where reports suggest that energy firms spend only about 0.2% of total revenue on cybersecurity-related measures [24]. It is well-accepted that initiatives like the CISA's Joint Cyber Defense Collaborative and NSA's Cybersecurity Collaboration Center are significant steps toward securing critical infrastructures. There should be a continuation of drills like the just concluded CyberStorm VIII with the inclusion of variety of partners to formulate scenarios for resilience and remediation strategies following massive cyberattacks. Besides, the continuous growth of research and development programs, such as DARPA's Rapid Attack Detection, Isolation, and Characterization Systems program, must be maintained to build tools and technologies that will prompt black start recoveries during a cyberattack on the power grid. An inclusion of the finance and banking sectors can also accelerate collaboration for research support from private sector, considering collobration with the energy service providers. Current achievements have been observed when both the public and private sectors display strong partnerships, as seen in the immediate action taken by Microsoft to upgrade its virus detection systems within three hours of being notified of the malicious malware named FoxBlade that severely hit Ukraine hours before the 2022 invasion [25]. Partnering with the Ukrainian administration, Microsoft aided the rapid dissemination of information to other EU states that could be significantly affected by such an event. Programs planned by investments from public resources should be held to train professionals in cybersecurity. There were great concerns about the unavailability of training for such professionals, which can be linked to the global cybersecurity workforce gap of nearly 2 million workers in 2022.

*4.3.6 Interoperable of cyber regulations for various future critical infrastructure*

The U.S. federal government has notified multiple critical infrastructure nodes of the intentions to regulate cybersecurity structures. After the Colonial Pipeline ransomware attack in 2021, the pipeline industry was regulated by The Transportation Security Administration (TSA). According to the operators of oil and gas pipelines, the regulations were '*full of awkward or inexplicable requirements that could endanger the safety of pipeline and supply of fuel*'. An all-encompassing solution cannot be developed given that various regulatory agencies supervise diverse sectors of critical infrastructure such as the energy, oil, gas, health care, and financial industries. This is more complicated since such new regulations must consider technical professionalism when being circulated. Simultaneously, industry officials who consider the cybersecurity of their operating and information technology systems as a future contemplation must ensure that their architecture is sufficiently provided with additional resources and developments.

Table 1. Standardization efforts to secure the energy infrastructures in the U.S. and Europe

| Regulation or law name | Review and major sections |
| --- | --- |
| **US Energy Policy Act** | Issues a directive to the FERC to authorize an Electric Reliability Organization (ERO) for setting up of compulsory security standards for the electricity network. |
| **US Energy Independence and Security Act** | Issues a directive to the 'National Institute of Standards and Technology (NIST) to for setting up essential cybersecurity standards for smart grids. |
| **US Executive Order 13636 "Improving Critical Infrastructure Cybersecurity"** | Commands the NIST to set up a cybersecurity structure permits for the protected smart grid development. The NIST Framework has been adopted by many US energy firms despite its voluntary and non-binding approach. |
| **US Presidential Policy Directive 21 "Critical Infrastructure Security and Resilience"** | Strengthens the regulation of federal agencies over important infrastructure like the energy sector. |
| **US Presidential Policy Directive 41"USA Cyber Incident Coordination"** | Strengthens the federal institutions assigned with the responsibility of developing national cybersecurity standards, such as FERC, NERC and the Department of Energy. |
| **US Executive Order 13800, "Strengthening the Cybersecurity of Federal Networks and Critical Infrastructure"** | Commands the Secretary of Energy to inquire with other federal agencies to evaluate the ability of energy utilities to resist cyberattacks through identifying key updates. |
| **EU Security of Supply Directive (SOS)** | Determines a set of security measures to protect the supply of EU's electricity and keep the internal electricity market functional, though without much of reference to cybersecurity. |
| **European Program for Critical Infrastructure Protection (EPCIP)** | Determines a typical inter-sectorial framework to ensure that critical infrastructures are secured and this encompasses cybersecurity, planned-out crime and natural calamities. |
| **European Critical Infrastructure Directive (ECI)** | Determine criteria for the identification and security of critical infrastructure and includes cybersecurity that is applicable to both the energy and transportation sectors. |
| **EU Security of Gas Supply Regulation** | Set up measures to ensure the security of the EU's gas supply. There is an adoption of a revised version in April 2017 that involves cybersecurity, to be enforced in the near future. |
| **EU Cyber Security Strategy** | Establish a set a of tactical priorities for cybersecurity of the EU energy infrastructure. |
| **EU Directive on the Security of Network and Information Systems (NIS)** | Affirms grounds to develop cybersecurity standards in the EU through the establishment of criteria for those who operate essential services and implement key policies for preventing and managing risks. |
| **Cybersecurity Package, proposed by the European Commission** | Includes recommendations to execute the NIS Directive. Also, the focus is given to strengthening the competencies of the ENISA to establish its permanent obligation and a system of European certification to construct an inner market for cybersecurity |

## 5. Conclusion

To conclude, the frequency and sophistication of malicious cyberattacks on energy systems increase as the world heads toward more digitalization. The potential for malicious actors to access and adversely affect physical energy assets via cyber means is a primary concern for countries around the globe. The Russian-Ukrainian conflict serves as an example of the growth of the utilization of cyberattacks on the energy infrastructure of a country to advance political, social, and military agendas. This work provides a discussion on the cybersecurity of the energy industry, and identifies the reasoning and motives behind cyberattacks. It also sheds the light on different standardization efforts made in this manner, and highlights a set of defensive steps to improve the resiliency of the grids against cyberattacks. It is recommended that future studies should analyze and investigate data, techniques, and procedures relative to cyberattacks in the energy field to assess the impact these attacks could have on nations since cyberattacks are currently being considered a weaponized tool in recent years. Finally, researchers need to explore methodologies that require bridging various disciplinary fields to understand better the challenges of securing the energy infrastructure, the most complex system humanity has ever built.